\documentstyle[12pt]{article}
\begin{document}
\renewcommand{\thefootnote}{\fnsymbol{footnote}}
\begin{titlepage}
\begin{flushright}
IJS-TP-16-94\\
TUM-T31-63-94 \\
June 1994 \\
\end{flushright}
\vspace{1cm}
\begin{center}
{\Large \bf Role of Scalar Meson Resonances
in $K_{L}^{0}\rightarrow\pi^{0} \gamma \gamma$ Decay} \\
\vspace{1cm}
{\large \bf S.Fajfer
\vspace{1cm}
} \\
\vspace{1cm}
{\it Institut "Jo\v zef Stefan", University of Ljubljana,
61111 Ljubljana, Slovenia}\\
and\\
{\it Physik Department, Technische Universit\"at M\"unchen\\
85748 Garching,
FRG}\\
\vspace{2cm}
\end{center}
\centerline{\large \bf ABSTRACT}
\vspace{0.5cm}
Corrections to $K_{L}^{0}\rightarrow\pi^{0} \gamma \gamma$ decay
induced by scalar meson exchange are studied within chiral perturbation theory.
In spite of bad knowledge of scalar-mesons parameters, the calculated branching
ratio can be changed by a few percent.

\end{titlepage}

\setlength {\baselineskip}{0.75truecm}
\parindent=3pt  % no indention of new paragraphs
\setcounter{footnote}{1}    % start footnotes at dagger instead
			     % of '*' (article style only)

\newcommand{\GeV}{\mbox{\enspace {\rm GeV}}}
\newcommand{\Tr}{\mbox{\rm Tr\space}}
\renewcommand{\thefootnote}{\arabic{footnote}}
\setcounter{footnote}{0}
\vspace{.5cm}
{\bf 1 Introduction}\\
\vspace{.5cm}

The $K\rightarrow\pi \gamma \gamma$ decays were subject of intensive
theoretical studies during last
few years \cite{EPRA,AP,EPRB,EPRC,EPRD,HS,CA,CEP}. The experimentally
measured branching ratio
\cite{NA31,E731}
of $K_{L}^{0}\rightarrow\pi^{0} \gamma \gamma$ is not so well
theoretically explained  as it seems from previous calculations.
In this decay,  ${\it O(p^{4})}$ terms in the chiral Lagrangian
 give the leading order contribution of Chiral Perturbation Theory (CHPT),
resulting in  the branching ratio $Br \simeq 0.7 \times 10^{-6}$. The amplitude
$K_{L}^{0}\rightarrow \pi \gamma \gamma$ is finite at one
loop level in CHPT. \\
The experimentally observed values are $(1.70\pm 0.3)
\times 10^{-6}$ ($NA31$ result) \cite{NA31}
and $(1.86\pm 0.60 \pm 0.60)) \times 10^{-6}$ (E731 result) \cite{E731}.
At the same time, the  observed invariant-mass distribution of the
final photons is
in good agreement with the theoretical predictions
\cite{EPRA,EPRB,EPRC,EPRD,CEP}.\\
The vector meson exchange, resulting in the ${\it O(p^{6})}$
contribution of CHPT,
was studied by authors of  \cite{EPRB,HS} and it was found that
this contribution is very important.
In addition to vector meson exchange present at
next-to-leading order, $\it O(p^{6})$
of CHPT, the two-pion intermediate state was taken into account \cite{CEP,CAM}.
It was found that these corrections raise the rate by $20\%$.
At the ${\cal O}(p^{4})$
order in the chiral Lagrangian, both vector and scalar resonance exchange
helps to explain $K\rightarrow \pi \pi \pi$  and $K\rightarrow \pi \pi$
amplitudes \cite{IP,EJW,SF1}.\\
Motivated by this effect, we investigate a role of the scalar resonances
in the $K_{L}^{0}\rightarrow\pi^{0} \gamma \gamma$ decay.
We notice, that scalar mesons also induce a contribution of
the ${\cal O}(p^{6})$
order in CHPT \cite{KMW,KM1,SF1,SF2}. Masses of $a_{0}(983)$
 and $f_{0}(975)$ scalar mesons
are close to the scale characterizing CHPT \cite{EG,EGR}
expansion $\Lambda \simeq 1 GeV$, and therefore they should be taken into
account. \\
The outline of the work is the following: in Sect. 2
we derive ${\cal O}(p^{6})$ effective Lagrangian for
$K_{L}^{0}\rightarrow\pi^{0} \gamma \gamma$ decay. In Sect. 3 we
discuss and comment our results.\\

{\bf 2 Scalar mesons and ${\cal O}(p^{6})$ effective Lagrangians}\\
\vspace{.5cm}

There are many attempts to understand the nature of scalar mesons
\cite{GP,GIK,J1,WI,MP,S}. In the chiral Lagrangian
we deal only with the quantum numbers of scalar mesons, and
we apply the approach of refs. \cite{EG,SF1,SF2}.\\
Very nice descriptions of CHPT up to ${\cal O}(p^{4})$  can be found in
ref. \cite{GL,EG,EGR,DGH}. We follow their notation. Here we describe only
a part of the chiral Lagrangian, necessary for our purpose.\\
The kinetic term of the Lagrangian describing scalar mesons is given by
\begin{eqnarray}
{\cal L}_{k}(S) & = &\frac{1}{2} {\rm tr} (\nabla_{\mu} S \nabla^{\mu} S
- M_{S}^{2} S^{2})\label{e1}
\end{eqnarray}
where $S$ is the scalar octet and $M_{S}$ corresponds to the scalar
masses in the chiral limit. For scalar singlet the kinetic term of
the Lagrangian is
\begin{eqnarray}
{\cal L}_{k}(S_{1}) &=& \frac{1}{2} (\partial^{\mu} S_{1} \partial_{\mu}
S_{1} - M_{S_{1}}^{2} S_{1}^{2})
\label{e2}
\end{eqnarray}
The scalar meson resonance  $f_{0}(983)$ can be
described as a linear combination of octet and singlet states of $SU(3)$,
while $a_{0}(975)$ completely belongs to its octet.\\
It means that we treat
$a_{0}$ and $f_{0}$ like $\rho$ and $\omega$ vector mesons, $a_{0}(975)$
is identified  with  $S_{3}$ and
\begin{eqnarray}
f_{0} (975) & = & \frac{1}{\sqrt{3}} S_{8} +\frac{2}{\sqrt{6}} S_{1}.
\label{e3}
\end{eqnarray}
Their interactions with Goldstone pseudoscalars can be described writing the
most general $SU(3)_{L} \times SU(3)_{R}$ Lagrangian taking into
account $C$ and $P$ properties of pseudoscalars and scalars \cite{GL,EG,EGR}
\begin{eqnarray}
{\cal L}_{SPP} & = & c_{d}  {\rm Tr} (S u_{\mu} u^{\mu}) + c_{m}  {\rm
Tr}( S {\chi}_{+}) + \bar{c}_{d}  S_{1} {\rm Tr} (u_{\mu} u^{\mu})
 + \bar{c}_{m}  S_{1} {\rm Tr}({\chi}_{+})\label{e4}
\end{eqnarray}
where
\begin{equation}
u_{\mu} = i u^{\dag} D_{\mu} U u^{\dag}\label{e9}\\
\end{equation}
\begin{eqnarray}
D_{\mu} U & = &\partial_{\mu} U + ie ({\rm A}_{\mu} U - U {\rm A}_{\mu}
)\label{e5}
\end{eqnarray}
\begin{eqnarray}
{\chi}_{+}& = &u^{\dag} \chi u^{\dag} + u {\chi}^{\dag} u \label{e6}
\end{eqnarray}
and $U = u^{2}$, is a unitary $3\times 3$ matrix, with $u = exp(-\frac {i}
{\sqrt{2}}
\frac{\Phi}{f})$, $\Phi = \frac{1}{\sqrt{2}}
\sum_{i=1}^{8}\lambda_{i}\varphi^{i}$, $Tr (\lambda_{i} \lambda_{j}) = \delta_
{ij}$,
$\Phi$ is the matrix of the
pseudoscalar fields,
${\rm A}_{\mu}$ is the electromagnetic field and  ${\chi}_{+}$ is related to
the quark mass matrix as in \cite{GL,EG}.\\
The experimental values of the decay widths of $f_{0}, a_{0}$ are given
in the Particle Data {\bf 92} \cite{PRP}.
\begin{equation}
{\Gamma}(f_{0}\rightarrow\pi\pi) = 36  {\rm MeV} \label{e10}
\end{equation}
\begin{equation}
{\Gamma}(a_{0}\rightarrow\eta\pi) = 59  {\rm MeV} \label{e11}
\end{equation}
Assuming that $M_{S} \simeq M_{S_{1}}$ and fitting the experimental
data for decay widths (\ref{e10}), (\ref{e11}) we derive
\begin{equation}
c_{d} =\pm 0.022 {\rm GeV},\enspace c_{m} = \pm 0.029 {\rm GeV}\label{e12}
\end{equation}
and
\begin{equation}
\bar{c}_{d} =  0.019 {\rm GeV},  \enspace \bar{c}_{m} = - 0.024 {\rm GeV}
\label{e13}
\end{equation}
or
\begin{equation}
\bar{c}_{d} =- 0.019 {\rm GeV},  \enspace \bar{c}_{m} = 0.011 {\rm GeV}
\label{e14}
\end{equation}
These fits are obtained from simultaneous fit of $l_{i}$,
$i = 1,6$ defined in \cite{EG} and $a_{0}$, $f_{0}$ decay widths (\ref{e13}),
(\ref{e14}), taking $\bar{c}_{d}$
 positive (\ref{e13}), and negative (\ref{e14}).
We take into account $\eta - {\eta}^{\prime}$
mixing through their mixing angle $\theta$. The results of our fit to $l_{i}$
are slightly different from those obtained in ref. \cite{EG}, where
$a_{0}$  decay rate (\ref{e11}) and large $N_{c}$ limit were used.
These authors \cite{EPRB,EG} emphasized that usual nonet assumption
for $\eta, \eta^{\prime}$ mesons which is widely used,
in the lowest order Lagrangian
${\cal L}_{2}$ is by no means unique.
In our proceeding calculations wee shall also use their fit for $c_{d}, c_{m},
{\bar c}_{m}, {\bar c}_{d}$.
\begin{equation}
c_{d} = \pm 0.032 {\rm GeV},\enspace c_{m} = \pm 0.042 {\rm GeV}\label{e12a}
\end{equation}
and
\begin{equation}
\bar{c}_{d} = \pm 0.019 {\rm GeV},  \enspace \bar{c}_{m} = \pm 0.024 {\rm GeV}
\label{e13a}
\end{equation}
We support the idea of scalar meson dominance in the counterterms
couplings, as well as in \cite{EG}.
In Table 1, we present both sets of the parameters.\\
In order to have two-photon-scalar couplings, we add to the
Lagrangian two photon
interaction with scalar mesons
\begin{eqnarray}
{\cal L}_{S\gamma \gamma} & = & g e^{2} Tr(Q^{2} S) F_{\mu \nu} F^{\mu \nu}
 + g^{\prime} e^{2} S_{1} F_{\mu \nu} F^{\mu \nu} \label{e15}
\end{eqnarray}
where $F_{\mu \nu}$ is the electromagenic field strenght tensor and $Q$
is the quark charge matrix. \\
We determine the constants  $g$ and $g^{\prime}$ by fitting
the exmerimental data \cite{PRP}. We are however, awere that
these fits should be taken with special caution.
\begin{eqnarray}
\Gamma (f_{0} \rightarrow 2 \gamma) & = & 0.56 \times 10^{-6} GeV\label{e16}
\end{eqnarray}
\begin{eqnarray}
\Gamma (a_{0} \rightarrow 2 \gamma) & = & 0.24 \times 10^{-6} GeV\label{e17}
\end{eqnarray}
We find following possible combinations for $g$ and $g^{\prime}$
\begin{eqnarray}
(1)\enspace g &=&0.07 GeV^{-1} \enspace g^{\prime} = 0.03 GeV^{-1}\label{e18}
\end{eqnarray}
\begin{eqnarray}
(2)\enspace g &=&0.07 GeV^{-1} \enspace g^{\prime} = -0.07 GeV^{-1}\label{e19}
\end{eqnarray}
\begin{eqnarray}
(3)\enspace g &=&-0.07 GeV^{-1} \enspace g^{\prime} = 0.07 GeV^{-1}\label{e18a}
\end{eqnarray}
\begin{eqnarray}
(4)\enspace g &=&-0.07 GeV^{-1} \enspace g^{\prime} = -0.03
GeV^{-1}\label{e19a}
\end{eqnarray}
Assuming nonet symmetry for scalar mesons  \cite{ABBC} requires
$c_{m} = {\bar c}_{m} = 0$. They use the
measured cross section $\sigma (\gamma \gamma \rightarrow \pi^{0} \eta) \simeq
30 nb$ at the $a_{0}(980)$  \cite{DST}.\\
The Lagrangian (4), (\ref{e15}) can be used \cite{EG,EJW}
to construct effective Lagrangian
describing two-pseudoscalar-two-photon couplings
dominated by scalar meson exchange.
Eliminating scalar mesons like we derive
\begin{eqnarray}
{\cal L}_{PP\gamma \gamma}^{s} & = &  g e^{2}
\frac{1}{M_{S}^{2}} F_{\mu \nu} F^{\mu \nu}
[c_{d} Tr(Q^{2} u^{\alpha} u_{\alpha})
- \frac{1}{3}c_{d} Tr(Q^{2}) Tr(u^{\alpha} u_{\alpha})\nonumber\\
&+& c_{m} Th( \chi_{+} Q^{2}) -c_{m}\frac{1}{3}
Tr( \chi_{+}) Tr (Q^{2})]\nonumber\\
& +&  g^{\prime} e^{2} \frac{1}{M_{S}^{2}} F_{\mu \nu} F^{\mu \nu}
 [{\bar c}_{d} Tr( u^{\alpha} u_{\alpha}) +
{\bar c}_{m} Tr(\chi_{+})]\label{e20}
\end{eqnarray}
The superscript $^{S}$ of ${\cal L}_{PP\gamma \gamma}$
is to show the presence of one strong vertex.
Accompanying this Lagrangian by the lowest order weak Lagrangian
\cite{EJW} describing
$K^{0} \rightarrow \pi^{0}, \eta, {\eta}^{\prime}$ transitions
\begin{equation}
{\cal L}_{w} = c_{2} Tr ({\lambda}_{6} u_{\mu} u^{\mu})
\label{e21}
\end{equation}
we easily obtain the amplitude, in which $\pi^{0}, \eta,
{\eta}^{\prime}$ are poles, as presented in fig. 1a. We take as in
\cite{EPRB},  $c_{2}/f^{4} = 9 \cdot 10^{-6} GeV^{-2}$ and
$f \simeq f_{\pi} = 0.933 GeV$.\\
{}From the direct weak kaon transition to pion and scalar meson, it is
possible to derive new contribution to
$K^{0} \rightarrow \pi^{0} \gamma \gamma$.
In this contribution a scalar meson decays into $2 \gamma$.\\
There are two procedures in the literature used to
determine effective weak Largangian:
the "factorization model"  \cite{EJW} and the "weak deformation model"
\cite{EPRA,EPRB,EPRC,EPRD,EJW}. It seems that
the "weak deformation model" has a realistic chance to describe
rather lagre number of processes accounting higher-order weak Lagrangian.
This model has obtained more confidence  after successful application
of the  ${\cal O}(p^{4})$ terms in  $K^{+} \rightarrow \pi^{+}
\gamma^{*}$ decay \cite{EPRB}, where weak counterterms
satisfy  scale independent
relations.
The weak Lagrangian containing vectors for
$K^{0} \rightarrow \pi^{0} \gamma \gamma$ was  derived
using this method \cite{EPRB}.
In order to maintain a consistent calculation of the vector
and scalar resonance
exchange, we apply this procedure, too. We find knowing (\ref{e20})
\begin{eqnarray}
{\cal L}_{PP\gamma \gamma}^{w} & = &  g e^{2} \frac{c_{2}
c_{d}}{M_{S}^{2}f^{2}} F_{\mu \nu} F^{\mu \nu}
[Tr(\lambda_{6} u^{\alpha} u_{\alpha}Q^{2})\nonumber\\
& - &\frac{4}{3}Tr(\lambda_{6} u^{\alpha} u_{\alpha}) Tr(Q^{2}) +
Tr(\lambda_{6}Q^{2} u^{\alpha} u_{\alpha})
 + Tr(\lambda_{6} u^{\alpha}Q^{2} u_{\alpha})]\nonumber\\
& +&
g^{\prime} e^{2} \frac{c_{2} {\bar c}_{d}}{M_{S}^{2}f^{2}}4 F_{\mu \nu} F^{\mu
\nu}
[Tr(\lambda_{6} u^{\alpha} u_{\alpha}Q^{2})]\label{e22}
\end{eqnarray}
where superscript $^{w}$ denotes the direct weak vertex (see fig. 1b.).\\
%\newpage

{\bf 3 Effective scalar meson contribution to the
decomposed $K_{L}^{0}\rightarrow\pi^{0} \gamma \gamma$ amplitude}\\
\vspace{.5cm}

The general decomposition for $K^{0} \rightarrow \pi^{0} \gamma \gamma$
amplitude is given by
\begin{eqnarray}
M(K^{0}(k) \rightarrow \pi^{0}(p) \gamma(q_{1}) \gamma(q_{2})) & = &
\epsilon_{\mu} (q_{1}) \epsilon_{\nu} (q_{2})[ \frac{A(y,z)}{m_{K}^{2}}
(q_{1}^{\nu} q_{2}^{\mu} - q_{1}\cdot q_{2} g^{\mu \nu})\nonumber\\
 +  2 \frac{B(y,z)}{m_{K}^{4}}(p \cdot q_{1} p\cdot q_{2} g^{\mu \nu}
+ q_{1}\cdot q_{2} p^{\mu} p^{\nu}& - & p\cdot q_{1} q_{2}^{\mu} p^{\nu} -
p\cdot q_{2} q_{1}^{\nu} p^{\mu})]\label{e23}
\end{eqnarray}
with dimensionless invariant amplitudes $A,B$ which are
functions of the Dalitz variables
\begin{eqnarray}
y & = & |k \cdot (q_{1} - q_{2})|/ m_{K}^{2}\label{e24a}
\end{eqnarray}
\begin{eqnarray}
z & = & (q_{1} + q_{2})^{2}/m_{K}^{2}\label{e24b}
\end{eqnarray}
Including loop effects at the order ${\cal O}(p^{4})$, vector mesons
\cite{EPRA,EPRB,EPRC,EPRD,HS,CA,CAM} and exchange of scalar
mesons calculated in this approach,
we can write
\begin{eqnarray}
A & = & \frac{ c_{2}}{f^{4}} \frac{m_{K}^{2} \alpha}{\pi}
[F(z/r_{\pi}^{2})(1 - \frac{r_{\pi}^{2}}{2} ) + F(z) (\frac{1 +r_{\pi}^{2}}{z}
-
1)\nonumber\\
&+& (a_{V} + a_{s}^{1})(3 -z + r_{\pi}^{2}) + a_{s}^{0}]
\label{e25}
\end{eqnarray}
where
\begin{eqnarray}
B & = & - 2 a_{V} \frac{c_{2}}{f^{4}} \frac{m_{K}^{2} \alpha}{\pi},
\enspace
r_{\pi} = \frac{m_{\pi}}{m_{K}}\label{e26}
\end{eqnarray}
\begin{eqnarray}
a_{V} &=& \frac{ 512 \pi^{2} h_{V}^{2} m_{K}^{2}}{9 m_{V}^{2}}\label{e27}
\end{eqnarray}
In \cite{EPRB} it was calculated that $a_{V} = -0.32$ without
$\eta - {\eta}^{\prime}$ mixing, and $a_{V} \simeq -0.19$ when
this mixing was included.
We find
\begin{eqnarray}
a_{s}^{1} & = & \frac{16 \pi^{2} m_{K}^{2} }{M_{S}^{2}}
[\frac{2 c_{d}g}{3} + \frac{2}{9} c_{d}g \beta(\theta) ]\label{e28}
\end{eqnarray}
with
\begin{eqnarray}
\beta({\theta}) & = & [-\frac{m_{K}^{2}}{m_{\eta}^{2}-
m_{K}^{2}} (cos \theta + 2{\sqrt 2} sin \theta)( cos \theta - {\sqrt 2} sin
\theta))\nonumber\\
& - & \frac{m_{K}^{2}}{m_{\eta}^{\prime 2}-
m_{K}^{2}} (sin \theta - 2{\sqrt 2} cos \theta)(sin \theta +{\sqrt 2} cos
\theta)]\label{e29}
\end{eqnarray}
where we take as usual $\theta \simeq 20^{o}$. We define
\begin{eqnarray}
a_{s}^{0} & = &  -2 a_{s}^{1} -\frac{16 \pi^{2}m_{K}^{2}}{M_{S}^{2}}
[\frac{4}{9} c_{m}g (1 + \beta(\theta)) + 4 g^{\prime} {\bar c}_{m}]\label{e30}
\end{eqnarray}
The scalar meson exchange does not influence $B$ invariant amplitude.
An interesting implication of this result is that
CP-conserving amplitude of
$K_{L}^{0} \rightarrow \pi^{0} e^{+} e^{-}{|}_{\gamma \gamma}$ decay,
proceeding
through $ \gamma \gamma $ states \cite{EPRA,AP,EPRB,BBB,DHV},
is not influenced by scalars.\\
As we have mentioned already, the choice of parameters
describing scalar mesons is the most troublesome part of this work.
We make all possible
allowed combinations of the parameters
$c_{d}, c_{m}, {\bar c}_{d}, {\bar c}_{m}, g$ and $g^{\prime}$
and we present the numerical results in Table 2, Table 3 and Table 4.
Without vector and scalar mesons the branching ratio was found to be
\cite{EPRA}
\begin{eqnarray}
Br_{0}(K_{L}^{0} \rightarrow \pi^{0}{\gamma \gamma}) & = &6.67 \times
10^{-7}\label{e31a}
\end{eqnarray}
When vector mesons and loops are included, the branching ratio is
\cite{EPRB,HS}
\begin{eqnarray}
Br(K_{L}^{0} \rightarrow \pi^{0}{\gamma \gamma}) & = & 8.87 \times
10^{-7}\label{e32}
\end{eqnarray}
for $a_{V}  =-0.32$, and for $a_{V} = -0.19$, when mixing with
$\eta - {\eta}^{\prime}$
was accounted is
\begin{eqnarray}
Br(K_{L}^{0} \rightarrow \pi^{0}{\gamma \gamma}) & = & 7.80
\times 10^{-7}\label{e32a}
\end{eqnarray}
As it can be seen from Table 2, 3, 4,
the largest contribution to the branching ratio is obtained
for the following choice of parameters
$g = \mp 0.07 GeV^{-1}$, $g^{\prime} = \pm 0.07 GeV^{-1}$,
 $c_{d} = \mp 0.032 GeV$
, $c_{m} = \mp 0.042 GeV$, ${\bar c}_{d} = \mp 0.019 GeV$,
${\bar c}_{m} = \mp0.024 GeV$ giving $a_{s}^{1} = -0.06$
and $a_{s}^{0}= 0.14$, where either upper or lower signs are taken
correspondingly. They give
\begin{eqnarray}
Br(K_{L}^{0} \rightarrow \pi^{0}{\gamma \gamma}) & = &
9.46 \times 10^{-7}\label{e33}
\end{eqnarray}
Taking values derived by \cite{ABBC} where
$gc_{d} = g^{\prime} {\bar c}_{d} \simeq \pm 0.16 \cdot 10^{-3}$,
 $a_{V} = -0.32$ we get
$Br(K_{L}^{0} \rightarrow \pi^{0}{\gamma \gamma})  =  8.83\times
10^{-7}$ for $+$ sign, while for $-$ sign it appears according \cite{ABBC},
$Br(K_{L}^{0} \rightarrow \pi^{0}{\gamma \gamma})  =  8.92 \times
10^{-7}$.
In the  calculation \cite{CEP}, where the exchange of two charged pions
was taken  into account, the physical amplitude
$K^{0}\rightarrow \pi^{0} \pi^{+} \pi^{-}$ was used to compute the
absorptive part of $K_{L}^{0} \rightarrow \pi^{0}{\gamma \gamma}$ amplitude
and then  subtracted dispersion relations were applied to obtain the
full amplitude. These corrections increase the branching ratio
by about $20\%$ in comparison with the
leading order term ${\cal O}(p^{4})$, created by pion and kaon loops.
In our analysis we do not add these corrections, since it was found
\cite{IP,EJW,SF1} that
the amplitude $K^{0}\rightarrow \pi^{0} \pi^{+} \pi^{-}$,
at ${\cal O}(p^{4})$  order is already
explained by the resonance exchange.\\
It was pointed out \cite{EPRB}, that there are many ${\cal O}(p^{8})$
contributions
related to vector mesons exchange, as well as some ${\cal O}(p^{5})$
 terms induced by $VP\gamma$ couplings which are not considered yet.
We do not take these  effects into account.\\
At ${\cal O}(p^{6})$ order of the weak
Lagrangian there are terms proportional to $l_{i}^{2}$,
induced  by ${\cal O}(p^{4})$ part of the chiral Lagrangian,
which could contribute, but their overall couplings are an order of
magnitude smaller than vector and scalar meson exchange considered in
the present paper.\\
Motiveted by CHPT study of \cite{EPRA}, $NA31$ \cite{NA31} has extracted the
bound on
$a_{V}$ from the Dalitz plot distribution of the two photon
$-0.32 < a_{V} < 0.19$.
{}From our result it is obvious that $a_{s}^{1}$ can increase
or decrease $a_{V}$ depending on the choice of the parameters
from$20\%$ to $30\%$.
On the other hand, only $a_{V}$ contributes to $B(y,z)$ invariant amplitude
of $K_{L}^{0} \rightarrow \pi^{0}{\gamma \gamma}$ decay.
In Table 2, 3, 4 we present the possible combinations of the parameters
$c_{d}, c_{m}, {\bar c}_{d}, {\bar c}_{m}, g, g^{\prime}$
and their influence on $Br$.\\
We see from results presented in Table 2 and 3
that the model used for $\eta - {\eta}^{\prime}$ mixing is
important, since if the nonet assumption for $\eta$'s is not used,
the branching ratio is increased by $15\%$.
In fig. 2, $\frac{1}{\Gamma} \frac{d\Gamma}{dz}$
is presented in the function of $\gamma \gamma$ invariant mass
for $a_{V}= a_{s}^{1} = a_{s}^{0}= 0$, for $a_{s}^{1} =
a_{s}^{0}=0$ $a_{V}=0.32$ and $a_{V}=0.32$, $a_{s}^{1} =-0.06$ and
$a_{s}^{0}= 0.14$.

Finally we can summarize:\\

(i) The corrections coming from scalar-meson exchange are rather small, but
not negligible. They might increase the branching ratio up to $6.7\%$.\\

(ii) The corrections strongly depend on the parameters determined by the scalar
meson data.\\

(iii) The CP-conserving $K_{L}^{0} \rightarrow \pi^{0} e^{+} e^{-}{|}_{\gamma
\gamma}$
decay rate is not influenced by scalar meson exchange.\\

(iv) The interference of vector and scalar mesons in the study of the
$A$ part of invariant amplitude in
$K_{L}^{0} \rightarrow \pi^{0}{\gamma \gamma}$ decay is not
negligible.\\

(v) It seems that the large experimental $Br(K_{L}^{0} \rightarrow
\pi^{0}{\gamma \gamma})$
can be theoretically explained when all possible contributions, coming from
loops at ${\cal O}(p^{4})$ order and the
accumulation of the smaller effects of the  ${\cal O}(p^{6})$,  or
even ${\cal O}(p^{8})$ order are taken into account, consistently.\\
%\vspace{1cm}
%
{\bf Acknowledgements}: The author thanks A.Buras  and B.Bajc
for useful discussions.\\

\vspace{2cm}

{\bf Figure Captions}\\

Fig. 1. Scalar meson exchange diagram for $K_{L}^{0} \rightarrow \pi ^{0}
\gamma \gamma$,
with $\pi, \eta, \eta^{\prime}$ poles (a), and with direct weak transition
(b).\\

Fig. 2. Normalized spectra in the ${\gamma \gamma}$ invariant mass
$z = (q_{1} + q_{2})^{2}/m_{K}^{2}$ for $a_{V} = 0$ (L - dotted curve),
$a_{V} = -0.32$, (V, L - dashed curve) and $a_{s}^{1} = -0.6$, $a_{s}^{0}=
0.14$
(S, V, L - full curve).\\

%\newpage
\begin{table}[h]
\begin{center}
\begin{tabular}{|c||c|c|c|c|c|c||c|}
\hline
$l_{i}$    &  V  & A & $S$  &$S_{1}$  &Total&Total[13]&
$l_{i}^{r}(M_{\rho})$\\
\hline\hline
$l_{1}$ & $0.6$ & $0$ &$-0.09$& $0.19$& $ 0.7$& $0.6$&$0.7\pm0.4$\\
\hline
$l_{2}$ & $1.2$ & $0$ &$0$& $0$& $ 1.2$& $1.2$& $ 1.3\pm0.7$\\
\hline
$l_{3}$ & $-3.6$ & $0$ &$0.27$& $0$& $ -3.33$& $-3.0$&$ -4.4\pm2.5$\\
\hline
$l_{4}$ & $0$ & $0$ &$-0.22$& $-0.48(-0.22)$& $-0.7(-0.44)$&$0.0$&
$0.3\pm0.5$\\
\hline
$l_{5}$ & $0$ & $0$ &$0.66$& $0$& $0.66$& $1.4$& $1.3\pm0.5$\\
\hline
$l_{6}$ & $0$ & $0$ &$-0.15$& $0.30(0.14)$& $-0.15(-0.01)$&$0.0$&
$-0.2\pm0.3$\\
\hline
$l_{8}$ & $0$ & $0$ &$0.45$& $0$& $0$&$0.9$& $0.9\pm0.3$\\
\hline
\end{tabular}
\end{center}
\caption{ $V, A, S, S_{1}$ contributions to the coupling constants $l_{i}^{r}$,
$i=1,2,3,4,5,6,8$ ($l_{7}$ is explained by the higher pseudoscalar meson state)
in units of $10^{-3}$ and compared with the values from [13].
The values are calculated for $c_{d}=\pm 0.022 GeV$ , $c_{m}=\pm 0.022 GvV$,
${\bar c}_{d} = 0.19 GeV$ and ${\bar c}_{m} =-0.024 GeV$ while in
the parentheses
are values obtained for ${\bar c}_{d} = -0.19 GeV$ and ${\bar c}_{m}
= 0.011GeV$. In the last column there are $l_{i}^{r}$ from [12], [13],
at the scale set to be equal $M_{\rho}$.}
\end{table}
\newpage
\begin{table}[h]
\begin{center}
\begin{tabular}{|r|r|r|r|r|r||c||r|}
\hline
$g$  & $g^{\prime}$ & $c_{d}$&$c_{m}$&
${\bar c}_{d}$&${\bar c}_{m}$&$ Br \cdot 10^{7}$&$\Delta Br [\%]$ \\
\hline\hline
$0.07$ & $0.03$ & $0.022$ & $0.029$ & $0.019$ & $-0.024$ & $8.54$&
$-3.7$ \\
\hline
$0.07$ & $0.03$ & $-0.022$ & $-0.029$ & $0.019$ & $-0.024$ & $9.06$& $2.1$\\
\hline
$0.07$ & $0.03$ & $0.022$ & $0.029$ & $-0.019$ & $0.011$ & $8.65$& $-2.5$ \\
\hline
$0.07$ & $0.03$ & $-0.022$ & $-0.029$ & $-0.019$ & $0.011$ & $9.18$&$3.5$ \\
\hline\hline
$0.07$ & $-0.07$ & $0.022$ & $0.029$ & $0.019$ & $-0.024$ & $8.80$&$-0.8$ \\
\hline
$0.07$ & $-0.07$ & $-0.022$ & $-0.029$ & $0.019$ & $-0.024$ & $9.33$&$5.2$ \\
\hline
$0.07$ & $-0.07$ & $0.022$ & $0.029$ & $-0.019$ & $0.011$ & $8.54$&$-3.7$ \\
\hline
$0.07$ & $-0.07$ & $-0.022$ & $-0.029$ & $-0.019$ & $0.011$ & $9.05$&$2.0$ \\
\hline\hline
$-0.07$ & $0.07$ & $0.022$ & $0.029$ & $0.019$ & $-0.024$ & $8.95$&$0.1$ \\
\hline
$-0.07$ & $0.07$ & $-0.022$ & $-0.029$ & $0.019$ & $-0.024$ & $8.44$&$-4.8$ \\
\hline
$-0.07$ & $0.07$ & $0.022$ & $0.029$ & $-0.019$ & $0.011$ & $9.23$&$4.1$ \\
\hline
$-0.07$ & $0.07$ & $-0.022$ & $-0.029$ & $-0.019$ & $0.011$ & $8.70$&$-0.8$ \\
\hline\hline
$-0.07$ & $-0.03$ & $0.022$ & $0.029$ & $0.019$ & $-0.024$ & $9.22$&$4.0$ \\
\hline
$-0.07$ & $-0.03$ & $-0.022$ & $-0.029$ & $0.019$ & $-0.024$ & $8.70$&$-2.0$ \\
\hline
$-0.07$ & $-0.03$ & $0.022$ & $0.029$ & $-0.019$ & $0.011$ & $9.10$&$2.6$ \\
\hline
$-0.07$ & $-0.03$ & $-0.022$ & $-0.029$ & $-0.019$ & $0.011$ & $8.58$& $-3.3$
\\
\hline
\end{tabular}
\end{center}
\caption{ The branching ratio $Br$ for $K_{L}^{0} \rightarrow \pi^{0}{\gamma
\gamma}$
decay.
For this calculation  all possible combinations of the
parameters $g$, $g^{\prime}$,
$c_{d}$, $c_{m}$, ${\bar c}_{d}$, ${\bar c}_{m}$ are used.
The $g$, $g^{\prime}$ are in $GeV^{-1}$ and  $c_{d}$, $c_{m}$, ${\bar c}_{d}$,
${\bar c}_{m}$
are given in $GeV$. In the last column the
 relative contribution of the scalar meson exchange is presented in percent.
$\Delta Br = (Br  - Br_{0})/Br_{0}$ with $Br_{0} = 8.87 \times 10^{-7}$
denotes the contribution without scalar mesons; $a_{V} = -0.32$ was taken.}
\end{table}
\newpage

\begin{table}[h]
\begin{center}
\begin{tabular}{|r|r|r|r|r|r||c||r|}
\hline
$g$  & $g^{\prime}$ & $c_{d}$&$c_{m}$&
${\bar c}_{d}$&${\bar c}_{m}$&$ Br \cdot 10^{7}$&$\Delta Br [\%]$ \\
\hline\hline
$0.07$ & $0.03$ & $0.022$ & $0.029$ & $0.019$ & $-0.024$ & $7.44$& $-4.6$ \\
\hline
$0.07$ & $0.03$ & $-0.022$ & $-0.029$ & $0.019$ & $-0.024$ & $8.04$& $3.1$\\
\hline
$0.07$ & $0.03$ & $0.022$ & $0.029$ & $-0.019$ & $0.011$ & $7.53$& $-3.5$ \\
\hline
$0.07$ & $0.03$ & $-0.022$ & $-0.029$ & $-0.019$ & $0.011$ & $8.15$&$4.5$ \\
\hline\hline
$0.07$ & $-0.07$ & $0.022$ & $0.029$ & $0.019$ & $-0.024$ & $7.65$&$-1.9$ \\
\hline
$0.07$ & $-0.07$ & $-0.022$ & $-0.029$ & $0.019$ & $-0.024$ & $8.29$&$6.3$ \\
\hline
$0.07$ & $-0.07$ & $0.022$ & $0.029$ & $-0.019$ & $0.011$ & $7.43$&$-4.7$ \\
\hline
$0.07$ & $-0.07$ & $-0.022$ & $-0.029$ & $-0.019$ & $0.011$ & $8.04$&$3.1$ \\
\hline\hline
$-0.07$ & $0.07$ & $0.022$ & $0.029$ & $0.019$ & $-0.024$ & $7.94$&$1.8$ \\
\hline
$-0.07$ & $0.07$ & $-0.022$ & $-0.029$ & $0.019$ & $-0.024$ & $7.35$&$-5.8$ \\
\hline
$-0.07$ & $0.07$ & $0.022$ & $0.029$ & $-0.019$ & $0.011$ & $8.19$&$5.0$ \\
\hline
$-0.07$ & $0.07$ & $-0.022$ & $-0.029$ & $-0.019$ & $0.011$ & $7.57$&$-3.0$ \\
\hline\hline
$-0.07$ & $-0.03$ & $0.022$ & $0.029$ & $0.019$ & $-0.024$ & $8.19$&$5.0$ \\
\hline
$-0.07$ & $-0.03$ & $-0.022$ & $-0.029$ & $0.019$ & $-0.024$ & $7.57$&$-3.0$ \\
\hline
$-0.07$ & $-0.03$ & $0.022$ & $0.029$ & $-0.019$ & $0.011$ & $8.08$&$3.4$ \\
\hline
$-0.07$ & $-0.03$ & $-0.022$ & $-0.029$ & $-0.019$ & $0.011$ & $7.47$& $-4.2$
\\
\hline
\end{tabular}
\end{center}
\caption{ The branching ratio $Br$ for $K_{L}^{0} \rightarrow \pi^{0}{\gamma
\gamma}$
decay.
For this calculation  all possible combinations of the parameters
$g$, $g^{\prime}$,
$c_{d}$, $c_{m}$, ${\bar c}_{d}$, ${\bar c}_{m}$ are used, as described in the
paper.
The $g$, $g^{\prime}$ are in $GeV^{-1}$ and  $c_{d}$, $c_{m}$,
${\bar c}_{d}$, ${\bar c}_{m}$
are given in $GeV$. In the last column the
 relative contribution of the scalar meson exchange is presented in percent.
$\Delta Br = (Br  - Br_{0})/Br_{0}$ with $Br_{0} = 7.80 \times 10^{-7}$
 denotes the contribution without scalar mesons. $\eta - {\eta}^{\prime}$
mixing
was taken into account and $a_{V} = -0.19$ was used.}
\end{table}
\newpage
\begin{table}[h]
\begin{center}
\begin{tabular}{|r|r|r|r|r|r||c||r|}
\hline
$g$  & $g^{\prime}$ & $c_{d}$&$c_{m}$&
${\bar c}_{d}$&${\bar c}_{m}$&$ Br \cdot 10^{7}$&$\Delta Br [\%]$ \\
\hline\hline
$0.07$ & $0.03$ & $0.032$ & $0.042$ & $0.019$ & $0.024$ & $9.18$& $3.5$ \\
\hline
$0.07$ & $0.03$ & $-0.032$ & $-0.042$ & $0.019$ & $0.024$ & $8.43$& $-5.0$\\
\hline
$0.07$ & $0.03$ & $-0.032$ & $-0.029$ & $-0.019$ & $-0.024$ & $8.58$& $3.2$ \\
\hline
$0.07$ & $0.03$ & $0.032$ & $0.042$ & $-0.019$ & $-0.024$ & $9.35$&$5.4$ \\
\hline\hline
$0.07$ & $-0.07$ & $0.032$ & $0.042$ & $0.019$ & $0.024$ & $8.33$&$-6.1$ \\
\hline
$0.07$ & $-0.07$ & $-0.032$ & $-0.042$ & $0.019$ & $0.024$ & $9.07$& $2.2$\\
\hline
$0.07$ & $-0.07$ & $-0.032$ & $-0.042$ & $-0.019$ & $-0.024$ & $9.46$& $6.7$ \\
\hline
$0.07$ & $-0.07$ & $0.032$ & $0.042$ & $-0.019$ & $-0.024$ & $8.68$&$-2.1$ \\
\hline\hline
$-0.07$ & $0.07$ & $0.032$ & $0.042$ & $0.019$ & $0.024$ & $9.46$&$6.7$ \\
\hline
$-0.07$ & $0.07$ & $-0.032$ & $-0.042$ & $0.019$ & $0.024$ & $8.68$&
$-2.1$\\
\hline
$-0.07$ & $0.07$ & $-0.032$ & $-0.042$ & $-0.019$ & $-0.024$ & $8.
33$& $-6.1$ \\
\hline
$-0.07$ & $0.07$ & $0.032$ & $0.042$ & $-0.019$ & $-0.024$ & $9.
07$&$2.2$ \\
\hline\hline
$-0.07$ & $-0.03$ & $0.032$ & $0.042$ & $0.019$ & $0.024$ & $8.58$&
$-3.2$ \\
\hline
$-0.07$ & $-0.03$ & $-0.032$ & $-0.042$ & $0.019$ & $0.024$ & $9.
35$& $5.4$\\
\hline
$-0.07$ & $-0.03$ & $-0.032$ & $-0.042$ & $-0.019$ & $-0.024$ & $9.
18$& $3.5$ \\
\hline
$-0.07$ & $-0.03$ & $0.032$ & $0.042$ & $-0.019$ & $-0.024$ & $8.43$&$5.0$ \\
\hline
\end{tabular}
\end{center}
\caption{ The branching ratio $Br$ for $K_{L}^{0} \rightarrow \pi^{0}{\gamma
\gamma}$
decay.
For this calculation  all possible combinations of the
parameters $g$, $g^{\prime}$,
$c_{d}$, $c_{m}$, ${\bar c}_{d}$, ${\bar c}_{m}$ are used.
The $g$, $g^{\prime}$ are in $GeV^{-1}$, while  $c_{d}$, $c_{m}$,
${\bar c}_{d}$, ${\bar c}_{m}$
are given in $GeV$ and their values are taken from [13].
In the last column the
relative contribution of the scalar meson exchange is presented in percent.
$\Delta Br = (Br  - Br_{0})/Br_{0}$ with $Br_{0} = 8.87 \times 10^{-7}$
 denotes the contribution without scalar mesons; $a_{V} = -0.32$ was taken.}
\end{table}

\end{document}